\documentclass[aps,pre,reprint,a4paper,showpacs,notitlepage,nofootinbib]{revtex4-1}

\usepackage{amsmath,amssymb}
\usepackage[dvipdfmx]{graphicx}
\usepackage{bm}
\usepackage{theorem}
\usepackage{grffile}

\makeatletter

\newcommand{\proofname}{\it{Proof}}
\makeatother

\usepackage{color}	

\begin{document}

\title{Fault Tolerance of Random Graphs with respect to Connectivity: \\Mean-field Approximation for Semi-dense Random Graphs}
\date{\today}

\author{Satoshi Takabe}
\email{s\_takabe@nitech.ac.jp}
\author{Takafumi Nakano}
\author{Tadashi Wadayama}
\affiliation{Department of Computer Science, Nagoya Institute of Technology, Gokiso-cho, Showa-ku, Nagoya, Aichi, 466-8555, Japan}

\begin{abstract}
The fault tolerance of random graphs with unbounded degrees with respect to connectivity is investigated, which relates to the reliability of wireless sensor networks with unreliable relay nodes.
The model evaluates the network breakdown probability that a graph is disconnected after stochastic node removal.
To establish a mean-field approximation for the model, we propose the cavity method for finite systems.
The analysis enables us to obtain an approximation formula for random graphs with
any number of nodes and an arbitrary degree distribution.
In addition, its asymptotic analysis reveals that
 the phase transition occurs in semi-dense random graphs whose average degree grows logarithmically.
These results, which are supported by numerical simulations, coincide with the mathematical results,
 indicating successful predictions by mean-field approximation for unbounded but not dense random graphs.

\end{abstract}


\maketitle
\emph{Introduction.}
Connectivity is a simple but fundamental property of graphs, as apparent from the fact that the first work on Erd\"os--R\'enyi (ER) random graphs examined the connectivity probability and
its critical phenomenon~\cite{ER59}.
Connectivity {is also regarded} as a measure of the reliability of systems {such as digital circuits~\cite{MS} and wireless communication networks}.
{In terms of wireless communication networks, one considers connectivity of a graph after} communication links, edges in a graph, are randomly removed.
Combined with random graph theory~\cite{Bol}, studies of reliability 
defined by connectivity~\cite{Ball} have provided theoretical insights into various technologies such as 
software fault tolerance~\cite{Avi} and the random key predistribution scheme~\cite{Key}. 

After the emergence of network science dealing with complex networks~\cite{comp1,comp2},
a fraction of a giant component (GC) has also been studied as a measure
 of network reliability against accidental fault of (or vulnerable attacks to) 
  nodes or edges~\cite{per1,per2,Cohen}.
{Its asymptotic property is related to the so-called percolation transition.}
The model has been applied to real networks such as power grids~\cite{pg1,pg2}.
Theoretically mean-field approximations 
contribute to those studies, 
revealing that network topology strongly affects fault-tolerance properties, e.g. a percolation threshold,
{which is proved generally later~\cite{Fou}.}

{The effectiveness of mean-field approximations for random graphs often depends on 
 the average degree.}
For dense graphs of $n$ nodes {with $\Theta(n)$ average degree},
 the replica method is often used to analyze disordered systems~\cite{SG}. 
{On the other hand,} many systems are {defined} on sparse graphs with bounded average degree: 
optimization problems~\cite{opt1,opt2}, community detection~\cite{com1,com2}, and so on.
The cavity method~\cite{cav0,cav1,cav2,cav3} and  generating function method~\cite{gf1,gf2} 
have been used {for mean-field approximations}. 
These methods provide analyses of properties and physical pictures 
 of those systems~\cite{sp1,sp2,sp3} {and 
mathematical conjectures such as percolation thresholds.}

{In the complex network analysis, \emph{semi-dense} random graphs are of interest because probabilistic analyses 
show the existence of phase transition phenomena of some crucial problems such as community detection~\cite{cd1,cd2}. 
Here, the semi-dense graph is a graph with $O(\ln n)$ average degree, which lies in the intermediate regime 
between dense and sparse cases.
In spite of its importance, mean-field approximations for sparse or dense random graphs are naively inapplicable
to semi-dense graphs because} the average degree is \emph{unbounded but not dense}.
Providing an example of mean-field analyses for disordered systems in the semi-dense regime will be 
useful not only to extend the scope of applications of the mean-field approximation but also
 to obtain a conjecture of a critical phenomenon.

The \emph{fault tolerance of random graphs with respect to connectivity} has attracted 
 interest along with development of IoT technologies~\cite{wsn1}.
In the model, disconnectivity probability of random graphs after stochastic node removals is evaluated, which
describes reliability of the wireless sensor network with unreliable relay nodes~\cite{The,Kuo}.
Nodes of this kind simplify, e.g., energy harvesting relay nodes~\cite{wsn2,Kaw}, which accidentally fails activation.
Although the asymptotic analysis has not been executed in general cases, the phase transition threshold  
is expected to lie in the $O(\ln n)$ regime~\cite{nnw}, which suggests that the model is a good example to examine the phase transition in {semi-dense graphs}.\

The motivation of this Letter is to establish a mean-field analysis for the node fault model {in semi-dense random graphs} and investigate its phase transition.
In the analysis, we first obtain a self-consistent cavity equation for disordered finite systems.
It enables us to obtain the approximation formula with finite $n$, which coincides asymptotically with
 numerical simulations.
The asymptotic analysis 
then reveals that the model exhibits a discontinuous phase transition with a threshold lying in the {semi-dense} regime.
Finally, we discuss the validity of the mean-field analysis and some future avenues of research.

\emph{Model Definition.}
Let $G=(V,E)$ be an undirected simple graph, i.e., a graph without self-loops and multiple edges, 
where $V=\{1,\dots,n\}$ and $E\subset V^2$ respectively represent a vertex set and an edge set.
Connectivity is defined as a graph property by which there exists a path between any pair of nodes in the graph.
We consider the node fault model~\cite{nnw}, also known as the random node breakdown~\cite{Cohen},
 in which each node is removed independently with constant probability $\epsilon$, called the node breakdown probability.
The resultant graph is named a survival graph.
The network breakdown probability $P^{(\epsilon)}(G)$ is defined as the {average} probability
 that a randomly generated survival graph is disconnected.
{The difference from the conventional percolation model is that a graph is at failure 
even when a node becomes isolated or a graph has more than two connected components.
In wireless sensor networks, this definition is reasonable because, in a strict sense, 
a network failure occurs when at least a relay node cannot transmit a packet to other nodes.}
 Our goal is to evaluate the network breakdown probability averaged over random graphs with unbounded degrees
and to investigate its phase transition.

{In this Letter, we consider a random graph ensemble $\Omega_n$ solely defined by}
the number of nodes $n$ and degree distribution $p_{n}(k)$ ($0\le k\le n-1$)~\cite{gf1}, which indicates that graphs sampled from $\Omega_n$ have no correlation such as degree--degree correlation~\cite{New1}, on average.
The network breakdown probability $P_{\Omega_n}(\epsilon)$ is then defined by the average of probability $P^{(\epsilon)}(G)$
 over a random graph ensemble $\Omega_n$.
Similar to the network breakdown probability, let us denote the average fraction of GC by $\rho_{\Omega_n}$.
It is noteworthy that the phase transition of $\rho_{\Omega_n}$ in the node fault model is known as network resilience~\cite{Cohen}.

\emph{Mean-field Analysis.}
Because connectivity is realized when all nodes in a survival graph belong to the GC,
we first evaluate the fraction of the GC $\rho_{\Omega_n}$.
The mean-field analysis of $\rho_{\Omega_n}$ has been investigated in~\cite{Buld} using the generating function method,
in~\cite{Shira,Wata,Zde} using the cavity method, and in~\cite{Non} using nonbacktracking expansions.
Herein, we briefly describe mean-field analysis based on the cavity method.

We introduce a classical spin model on a given graph $G=(V,E)$ as in~\cite{Wata}.
We define a binary variable $s_i$ for the $i$-th node by $s_i=1$ if the node is active and $s_i=0$ otherwise. 
The other binary variable $\sigma_i$ is introduced to describe whether the $i$-th node is contained to the GC:
$\sigma_i=1$ if the node does not belong to the GC, and $\sigma_i=0$ otherwise.
After some nodes have broken down, i.e., the sequence $\{s_i\}$ is fixed, a constraint for the $i$-th node to belong to the GC
is given as
\begin{equation}
\sigma_i= \prod_{j\in\partial i}(1-s_j+s_j\sigma_j), \label{eq_appb1}
\end{equation}
where $\partial i$ represents the set of neighbors of the $i$-th node.
To find the GC satisfying Eq.~(\ref{eq_appb1}), we use the sum-product algorithm~\cite{cav2}.
Given that $G$ is a tree, $\sigma_i$ can be obtained exactly as
\begin{equation}
\sigma_i= \prod_{j\in\partial i}(1-s_j+s_jm_{j\rightarrow i}), \label{eq_appb1a}
\end{equation}
where $m_{j\rightarrow i}$ is a message from the $j$-th node to the $i$-th node, which takes a value of one if $\sigma_j=1$ holds
 on the cavity graph $G\backslash \{i\}$, and zero otherwise.
These messages satisfy cavity equations for the GC~\cite{Wata}, expressed as
\begin{equation}
m_{i\rightarrow j}= \prod_{k\in\partial i\backslash \{j\}}(1-s_k+s_km_{k\rightarrow i}). \label{eq_appb1b}
\end{equation}
Although the sum-product algorithm is no longer exact for graphs with cycles, it is regarded as the Bethe--Peierls approximation~\cite{Bet,KS}.

To calculate the random graph average, the replica symmetric (RS) cavity method is applied to the system.
Although earlier works directly take the large-$n$ limit, as with the case of the conventional cavity method,
the method for random graphs with \emph{finite} $n$ is proposed here
 to examine graphs with unbounded average degrees.

Considering that the messages $\{m_{i\rightarrow j}\}$ are updated recursively by the cavity equation~(\ref{eq_appb1b}),
let $I^{(t)}_n$ be the probability that the message $m_{i\rightarrow j}$ takes one for randomly chosen graph $G$ 
 and its edge $(i,j)$ from $\Omega_n$ at the $t$-th iteration step.
If the correlation between any pair of spins is negligible,
then $I_n^{(t)}$ satisfies the following relations:
\begin{equation}
I^{(t+1)}_n=\sum_{k=1}^n \frac{kp_n(k)}{\langle k\rangle}(\epsilon+(1-\epsilon)I^{(t)}_n)^{k-1}, \label{eq_appb5}
\end{equation}
where $\langle k\rangle\equiv \sum_{k=1}^nkp_n(k)$ represents the average degree of random graphs.
A fixed-point solution $I_n$ is obtained in the large-$t$ limit, which is given as
\begin{equation}
I_n=\sum_{k=1}^n \frac{kp_n(k)}{\langle k\rangle}(\epsilon+(1-\epsilon)I_n)^{k-1}. \label{eq_appb5a}
\end{equation}
The average fraction of the GC under the mean-field approximation is then obtained as
\begin{equation}
{\rho^{\mathrm{MF}}_{\Omega_n}=(1-\epsilon)\left\{1-\sum_{k=1}^n p_n(k)[\epsilon+(1-\epsilon)I_n]^{k}\right\}.} \label{eq_appb5b}
\end{equation}

The breakdown probability $P_{\Omega_n}(\epsilon)$ is then approximated using the evaluation presented above.
As a graph is connected when each active node belongs to the GC of the survival graph,
the approximation formula of the network breakdown probability is given as
\begin{align}
P_{\Omega_n}^{\mathrm{MF}}(\epsilon)&=1-\left(\epsilon+\rho^{\mathrm{MF}}_{\Omega_n}\right)^n\nonumber\\
&=1-\left(1-(1-\epsilon)\sum_{k=0}^n p_n(k)[\epsilon+(1-\epsilon)I_n]^k\right)^n. \label{eq_c5}
\end{align}

Equation~(\ref{eq_appb5a}) has a trivial solution $I_n=1$ corresponding to a ``broken phase'' in which $\rho^{\mathrm{MF}}_{\Omega_n}=0$ holds.
It also possibly has another solution $\tilde{I}_n$ in $[0,1)$. 
Because $\epsilon$ is assumed to be constant, $P_{\Omega_n}^{\mathrm{MF}}(\epsilon)$ converges to one if $\tilde{I}_n$ remains positive {as $n\rightarrow \infty$.}
We therefore examine the case in which the average degree and resultant $\tilde{I}_n$ depend on $n$
 using the RS cavity method for finite-size systems.

\emph{{Asymptotic Analysis and Finite-Size Effects}.}
To investigate the phase transition of connectivity, here we {conduct} an asymptotic analysis of the approximation formula.
Unfortunately, it is difficult to obtain an explicit form of the critical threshold in general.
Alternatively, we describe the analytical results for three well-known random graph ensembles.
{In addition, we examine the finite-size effects of the approximation by comparing it with numerical simulations
in some cases.}

\emph{(i) ER random graphs.}
The ER random graph with $n$ nodes is characterized by the binomial degree distribution, 
$p_n(k)=\binom{n-1}{k}(1-p)^{n-k-1}p^k$,
where $p\equiv \langle k\rangle/(n-1)$ and $\langle k\rangle$ is the average degree.
Then Eq.~(\ref{eq_appb5a}) reads as
\begin{equation}
I_n=\left[1-p(1-\epsilon)(1-I_n)\right]^{n-1}. \label{eq_er1}
\end{equation}
The approximate network breakdown probability $P_{n, \langle k\rangle}^{\mathrm{MF}}(\epsilon)$ is given as
\begin{equation}
P_{n,\langle k\rangle}^{\mathrm{MF}}(\epsilon)=1-\left[1-(1-\epsilon)\left\{1-p(1-\epsilon)(1-I_n)\right\}^{n-1}\right]^n. \label{eq_er2}
\end{equation}
{If we set $\langle k\rangle=d\ln n$ with a positive constant $d$ and assume that the leading term of  $\tilde{I}_n$ is given by  $an^{-l}$ ($a,l>0$), Eq.~(\ref{eq_er1}) reads
\begin{align}
\tilde{I}_n&=\left\{1-\frac{d\ln n}{n-1}(1-\epsilon) \left[1-an^{-l}+o(n^{-l})\right]\right\}^{n-1}\nonumber\\
&=n^{-{d(1-\epsilon)}}(1+O(\max\{n^{-l}\ln n,n^{-1}\ln^2n\})). \label{eq_d3}
\end{align}
This relation indicates that the asymptotic form is given by $\tilde{I}_n=n^{-{d(1-\epsilon)}}(1+o(1))$.}

This asymptotic form is then substituted to $I_n$ in Eq.~(\ref{eq_er2}).
Using the fact that, in the large-$n$ limit,
  $(1-an^{-b})^n$ converges to zero when $b<1$, to $e^{-a}$ when $b=1$ and to one otherwise,
  the threshold $\epsilon^\ast$ of node breaking probability is obtained by $\epsilon^\ast=1-1/d$.
In other words, $P_{n,\langle k\rangle}^{\mathrm{MF}}(\epsilon)$ jumps from zero to one at the threshold $\epsilon^\ast$.
This result suggests that ER random graphs loses connectivity even when $\epsilon=0$ if $\langle k\rangle<\ln n$ holds,
which is proved in~\cite{ER59}. 

\emph{(ii) Regular random graphs.}
The regular random (RR) graph is characterized by $p_n(k)=\delta_{k,\langle k\rangle}$ where $\delta_{n,m}$ represents Kronecker's delta.
Although RR graphs themselves have connectivity with high probability if $\langle k\rangle\ge 3$~\cite{Wor},
results show that connectivity is lost by node breakdowns for any $\epsilon> 0$.
As in the case of ER random graphs, we scale the average degree as $\langle k\rangle =d\ln n$.
The asymptotic analyses yield that the leading term of $\tilde{I}_n$ is given as $n^{-d\ln(1/\epsilon)}$.
The threshold is given by $\epsilon^\ast=e^{-1/d}$.

\begin{figure}[]
\begin{center}
\includegraphics[width=0.95\linewidth]{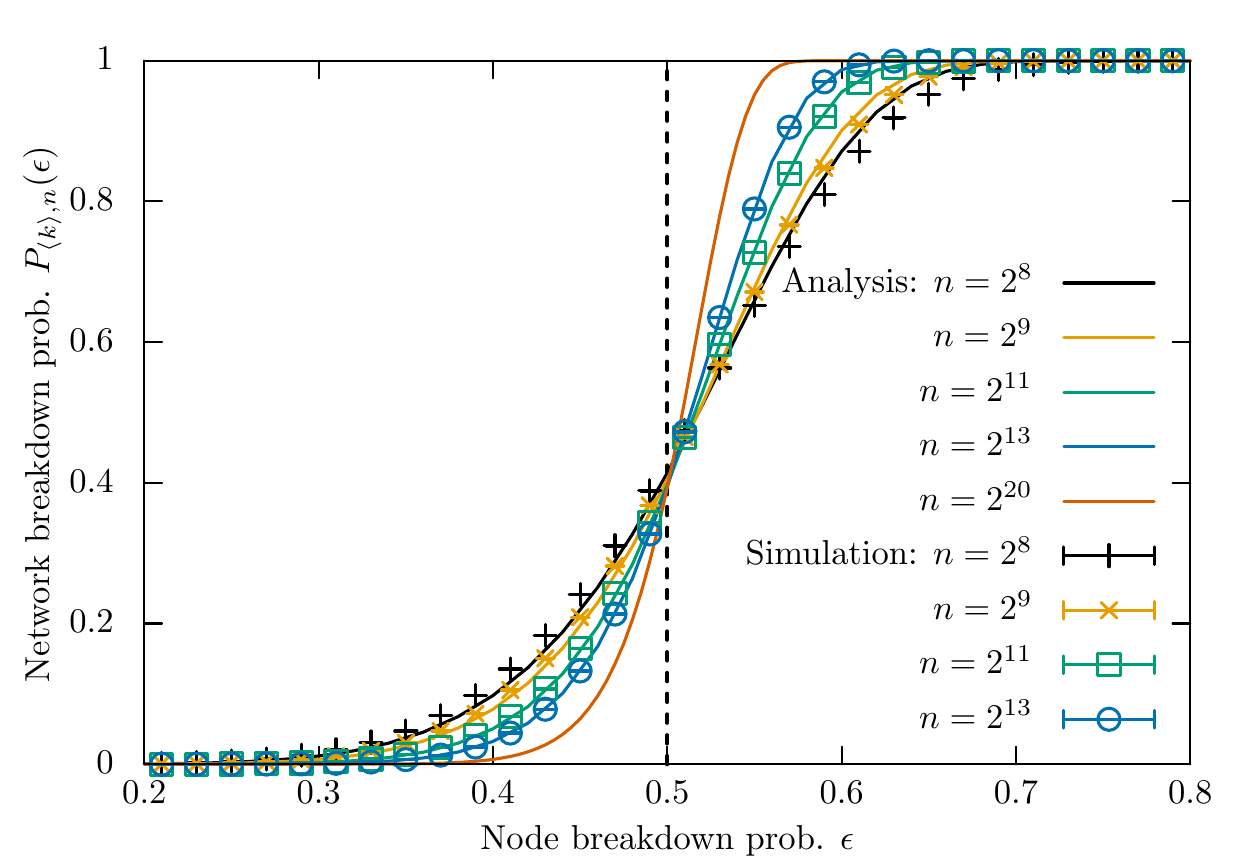}
\end{center}
\caption{Network breakdown probability $P_{\langle k\rangle,n}(\epsilon)$ as a function of node breakdown probability $\epsilon$ for regular random graphs with $n$ nodes of which the degree is $\langle k\rangle=\log_2n$.
Solid lines represent evaluations by the mean-field approximation.
Symbols show numerical results averaged over $10^6$ randomly generated survival graphs.
The vertical dashed line represents the phase transition threshold $\epsilon^\ast=1/2$ by asymptotic analysis.
}\label{zu_2}
\end{figure}

Figure~\ref{zu_2} presents numerical results and mean-field evaluations of the network breakdown probability
for RR graphs with degree $\langle k\rangle =\log_2n$.
The numerical simulation is performed by generating $10^3$ random graphs and by checking the connectivity of $10^3$ survival graphs per random graph using depth-first search.
Results show that the mean-field evaluations by Eqs.~(\ref{eq_appb5a}) and (\ref{eq_c5}) are
 asymptotically close to the numerical results, indicating that the analytical result $\epsilon^\ast=1/2$ is correct.
{Moreover, the mean-field approximation well predicts the network breakdown probability for sufficiently large $n$,
indicating that the finite-size effect is relatively small.}

\emph{(iii) Scale-free networks.}
We also consider scale-free (SF) networks whose degree distribution is given by a power-law distribution $p_n(k)\propto k^{-\gamma}$~\cite{comp2}.
To parameterize the ensemble with the scaling factor $\gamma$ and average degree $\langle k\rangle$, we consider random SF networks with
the minimum degree $\tilde{k}$. The degree distribution is therefore written as 
\begin{equation}
p_n(k)=C_n^{-1}k^{-\gamma}\, (\tilde{k}\le k< n),\, C_n=\zeta(\gamma,\tilde{k})-\zeta(\gamma,n), \label{eq_sf1}
\end{equation}
where $\zeta(s,a)=\sum_{k=0}^{\infty}(k+a)^{-s}$ is the Hurwitz zeta function.
Then, the average degree is given as $\langle k\rangle=C_n^{-1}[\zeta(\gamma-1,\tilde{k})-\zeta(\gamma-1,n)]$, which is well-defined when $\gamma>2$.
As the above cases illustrate, it is apparent that the possible threshold is trivial, i.e., $\epsilon^\ast =0$ when $\langle k\rangle=O(1)$.
If the average degree is again scaled as $d\ln n$, then the minimum degree also depends on $n$ because $\langle k\rangle=(\gamma-1)/(\gamma-2)\tilde{k}+O(1)$ holds.
The self-consistent equation of $I_n$, Eq.~(\ref{eq_appb5a}), is expressed as
\begin{equation}
I_n=\frac{[\epsilon+(1-\epsilon)I_n]^{\tilde{k}-1}}{C_n\langle k\rangle}\Phi\left(\epsilon+(1-\epsilon)I_n,\gamma-1,\tilde{k}\right), \label{eq_sf2}
\end{equation}
where $\Phi(z,s,a)=\sum_{k=0}^\infty z^k/(k+a)^{s}$ is called the Lerch transcendent.

\begin{figure}[]
\begin{center}
\includegraphics[width=0.95\linewidth]{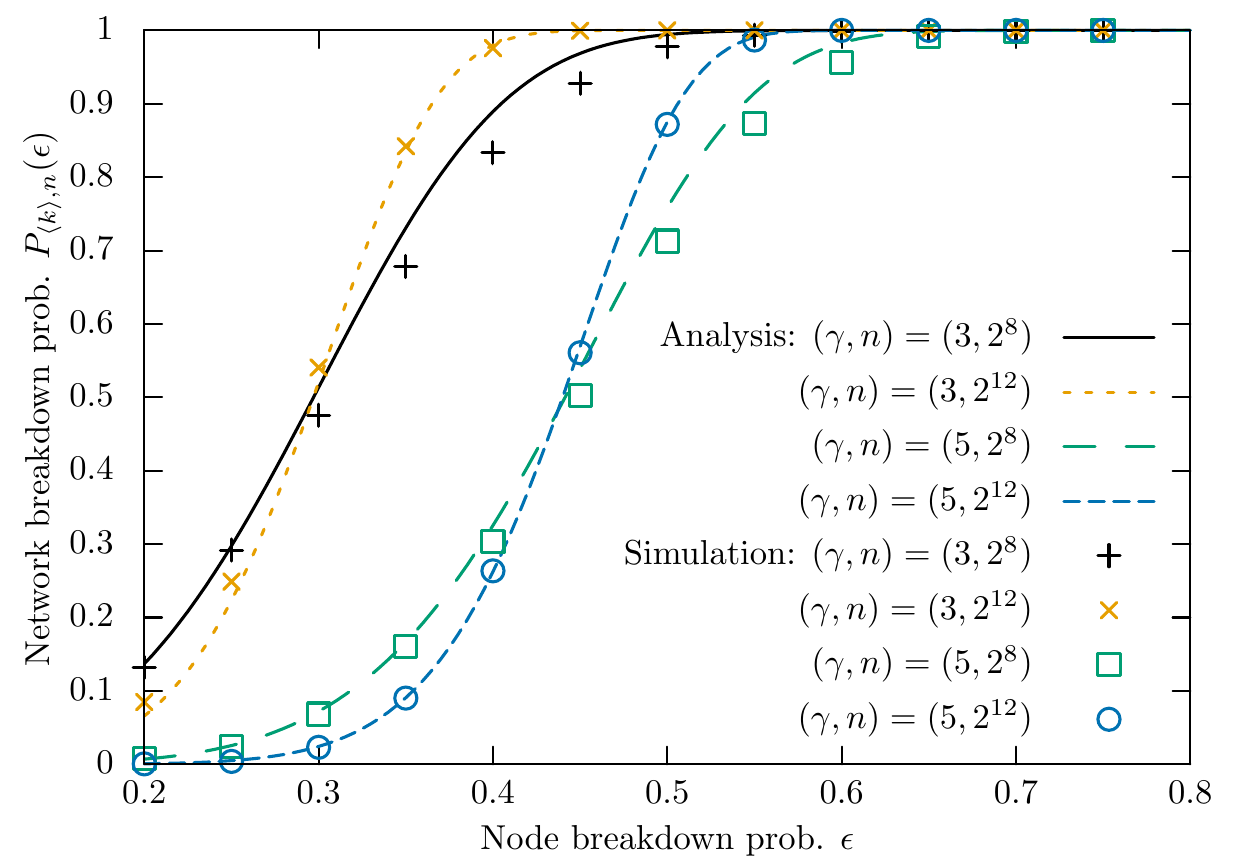}
\end{center}
\caption{{Network breakdown probability $P_{\langle k\rangle,n}(\epsilon)$ as a function of node breakdown probability 
$\epsilon$ for scale-free networks with $n$ nodes whose scaling factor is $\gamma$ and minimum degree is $\tilde{k}=(\gamma-2)/(\gamma-1)\log_2n$.
Lines represent evaluations by the mean-field approximation.
Symbols show numerical results averaged over $10^4$ randomly generated survival graphs.
}}\label{zu_2a}
\end{figure}

{
To check the accuracy of the mean-field theory, 
we compare it with numerical simulations in Fig.~\ref{zu_2a}.
The minimum degree $\tilde k$ is fixed to $(\gamma-2)/(\gamma-1)\log_2n$ to scale 
the average degree as $\log_2 n$.
It is found that the finite-size effect seems sufficiently small even in scale-free networks.
Although the scale-free network generally has relatively large finite-size effects due to 
its power-law degree distribution and the existence of  a hub, the result suggests that
the mean-field approximation is reasonably effective for analyzing the semi-dense SF network. 
}

{Like other random graphs, asymptotic analysis is also applicable to the random SF networks.}
The leading term of $\tilde{I}_n$ is found to be $n^{-\tilde{d}\ln(1/\epsilon)}$ 
by an asymptotic form of $\Phi(z,s,a)$~\cite{HL}, and 
 the threshold is given as $\epsilon^\ast=e^{-(\gamma-1)/[(\gamma-2)d]}$.

\begin{figure}[]
\begin{center}
\includegraphics[width=0.95\linewidth]{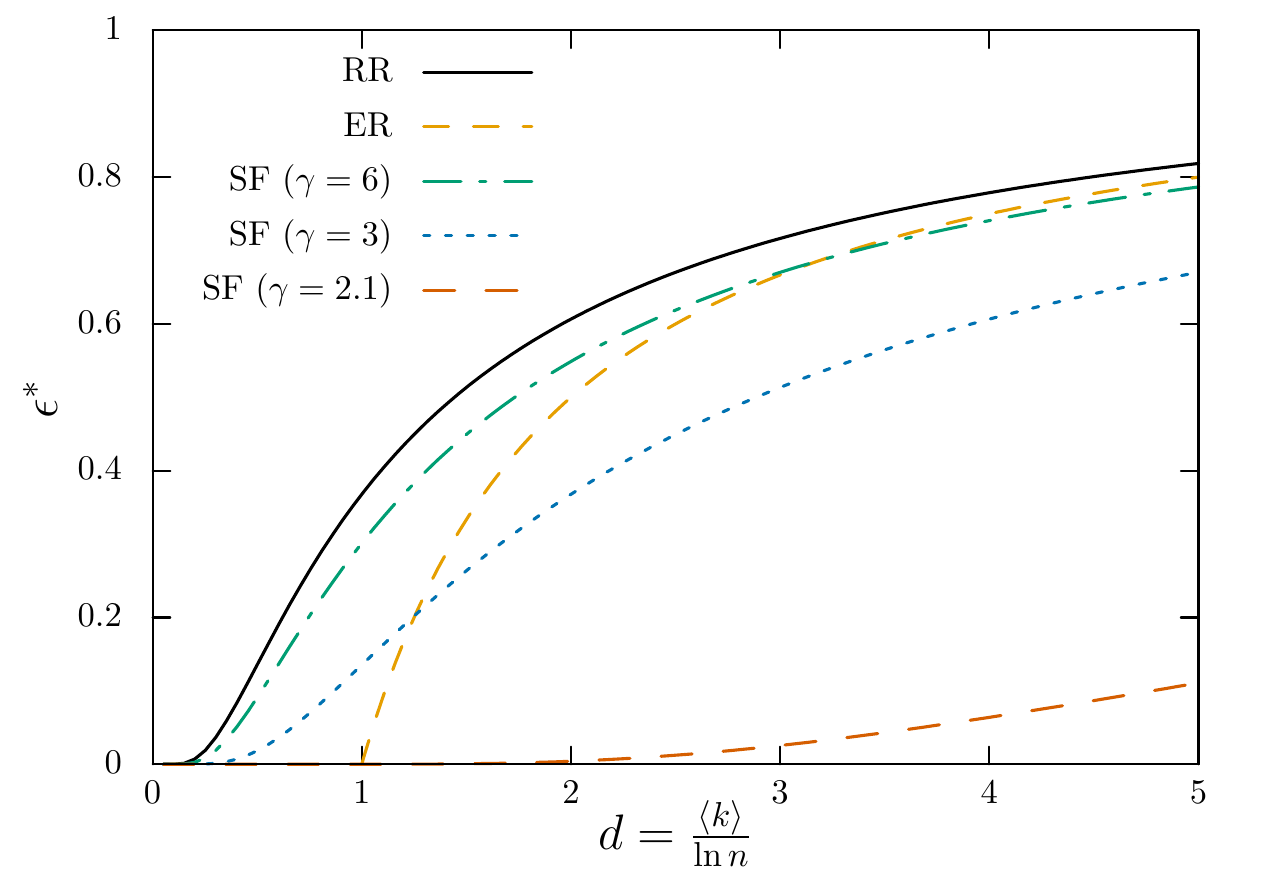}
\end{center}
\caption{Phase transition threshold $\epsilon^\ast$ as a function of a factor of the average degree $\langle k\rangle$ given as $d=\langle k\rangle/\ln n$.
}\label{zu_3}
\end{figure}

{\emph{Comparison of Reliability.}} 
The asymptotic analysis enables us to compare the reliability of each random graph ensemble.
Figure~\ref{zu_3} presents a summary of the thresholds $\epsilon^\ast$ of random graphs of three types as a function of $d\equiv \langle k\rangle/\ln n$.
Although the threshold for ER random graphs exists only when $d\ge 1$, other ensembles have the threshold for any $d$.
With factor $d$ fixed, the threshold of SF networks increases monotonously as the power $\gamma$ increases.
Particularly, $\epsilon^\ast$ for RR graphs coincides with that for SF network in the large-$\gamma$ limit,
  which is always larger than that for ER random graphs.
This fact is intuitive because the minimum degree, which is maximized by eliminating the variance of degree distribution,
  influences connectivity.
We can thereby conclude that RR graphs is the most reliable ensemble among them 
 in the node fault model when average degrees are fixed.

\emph{Summary and Discussion.}
In this Letter, we presented consideration of the node fault model defined by graph connectivity.
The mean-field analysis yields the approximation formula available for any number of nodes $n$ and an arbitrary degree distribution.
It is also revealed that the model exhibits a discontinuous phase transition in {semi-dense random graphs whose} 
 average degree is given as $O(\ln n)$.
Below the threshold $\epsilon^\ast$ {of the node breakdown probability}, 
the network breakdown probability vanishes in the large-system limit,
although it converges to one when $\epsilon>\epsilon^\ast$.
The accuracy of our approximation formula and thresholds are confirmed by
numerical simulations. 

The mathematical rigor of our analytical results is worth noting.
For ER random graphs, in fact, we can prove the correctness of our prediction~\cite{sw} by extending the result by Erd\"os and R\'enyi~\cite{ER59}.
In contrast, because the proof is specific to the ensemble, the rigor of our predictions on other ensembles is left as an open problem.
We emphasize, however, that the correctness of the specific case is strong evidence of the validity of our analyses.

{The validity of the approximation itself is also a crucial problem.}
In sparse random graphs, the use of approximation is justified by the \emph{locally tree-like structure} of graphs
that the length of cycles are typically scaled logarithmically~\cite{sp3}.
In this manner, our analysis is justified because the length of cycles is evaluated as $O(\ln n/\ln\ln n)$~\cite{comp1}.
From a mathematical perspective, however, the justification remains an open problem
 because the known rigorous results~\cite{DM} only treat random graphs with bounded degrees.
In other words, our results provide not only conjectures related to analytical results
  but also a challenging open problem in the mean-field theory.

The node fault model presented in this Letter has some possible extensions.
The first direction is consideration of intentional attacks to nodes and correlations in graphs.
These are analogous to the original node fault model with respect to percolation~\cite{Coh2,Vaz}.
Studies of these extensions reveal the network topology dependence on reliability.
It is another direction to examine the node fault model on random graphs in metric space.
The representative model is known as a random geometric graph~\cite{Gil}, which is often used as an abstract model
 of wireless ad-hoc {networks~\cite{San,Det,ad1,ad2}}.
These extensions enable us to optimize networks while conserving reliability, as in the case of percolation~\cite{Pau}.


\begin{acknowledgments}
The authors warmly thank T. Nozaki, T. Hasegawa, K. Hukushima, and Y. Kabashima for fruitful discussions.
ST also thank T. Kawamoto for carefully reading the manuscript.
This research is partially supported by JSPS KAKENHI Grant Nos. 16H02878, 16K14267 (TW), 17H06758, and 19K14613 (ST).
\end{acknowledgments}

\end{document}